\documentstyle[aaspp4]{article}
\newcommand{\rightleftarrow}{\mbox{
\raisebox{-.7ex}
{$\stackrel{\textstyle \rightarrow}{\textstyle \leftarrow}$}}}

\begin{document}


\title{\bf PRIMORDIAL FRACTAL DENSITY PERTURBATIONS AND  STRUCTURE
FORMATION IN THE UNIVERSE : 1-DIMENSIONAL COLLISIONLESS SHEET MODEL}

\author{Takayuki {\sc Tatekawa}$^1$ and Kei-ichi {\sc Maeda}$^{1,2}$}

\affil{1 Department of Physics, Waseda University, 3--4--1 Okubo,
Shinjuku, Tokyo 169-8555, JAPAN\\2  Advanced Research Institute for
Science and Engineering, Waseda University, Shinjuku, Tokyo 169-8555,
JAPAN
\\ E-mail: tatekawa,
maeda@gravity.phys.waseda.ac.jp}

\begin{abstract}
Two-point correlation function of galaxy distribution shows that the
structure in the present Universe is scale-free up to a certain scale (at
least several tens Mpc), which suggests that a fractal structure may exist. 
If small primordial density fluctuations have a fractal structure,   the 
present
fractal-like nonlinear structure below the horizon scale could be
naturally explained. We analyze the time evolution of fractal density 
perturbations in
Einstein-de Sitter universe, and study how the perturbation evolves and
what kind of nonlinear structure will come out. We assume a one-dimensional
collisionless sheet model with initial Cantor-type fractal perturbations.
The nonlinear structure seems to approach some attractor with a
unique fractal dimension, which is independent of the fractal
dimensions of initial perturbations. A discrete self-similarity in the phase
space is also found when the universal  nonlinear fractal structure is reached.
\end{abstract}

{\it Subject headings}:
cosmology: theory --- large-scale structure --- fractal ---
1-dimensional sheet model




\section{INTRODUCTION}
The present Universe shows a variety of structures.  The
galaxies are not distributed randomly in the Universe.
  Totsuji \& Kihara (1969) and Peebles (1974) showed that the observed two-point
correlation function
$\xi(r)$ is given by a power law
with respect to a distance
$r$ as $\xi(r) \sim r^{-\gamma}$ with $\gamma \sim 1.8$.
  The recent galaxy surveys also agree with this result, i.e. the
power $\gamma$ is nearly equal to $1.8$ (CfA (Geller \& Hachra (1989)),
LCRS (Jing, Mo, \& B\"{o}rner (1998)), and ESP (Guzzo et al. (1998, 1999)).
This may imply that the present distribution
of galaxies is fractal.
Sylos Labini, Montuori, \& Pietronero (1998) have also claimed that all 
available data
are consistent with a fractal structure with the dimension $D
\sim 2$ up to the deepest observed scale ($1000 h^{-1}$[Mpc]).
However, the observation of Cosmic
Microwave Background Radiation (CMBR) has revealed that the Universe
in the recombination era is homogeneous and isotropic at least in
very large scale. Although CMBR observation seems to be more
reliable, we should not decide yet whether the large scale
structure of the Universe is really fractal up to the horizon
scale or not. To answer this question more definitely, we should await 
forthcoming next galaxy survey projects (Colless (1995), Maddox (1997), 
Loveday \& Pier
(1998), Knapp et al. (1999)).

However, since it seems true that the galaxy distribution is
really fractal up
to a certain scale, one may ask how such a structure is formed
in the evolution of the Universe. One of the most plausible
explanations is that the nonlinear dynamics of the perturbations
will provide such a scale-free structure during the evolution of
the Universe. The pioneering work to explain the power-law
behavior in nonlinear stage has been done by Davis
\& Peebles (1977).
They assume a self-similar evolution of density fluctuation and
some additional condition, i.e. a physical velocity $\dot{r}$
vanishes in nonlinear regime. Then they showed a relation
between the power index
$\gamma$ of two-point correlation function and that of initial
power spectrum
$n$ as $\gamma=3(n+3)/(n+5)$. If we have
$n=0$, then we find that $\gamma =1.8$.
Since their additional condition is not trivial and might not be
appropriate, Padmanabhan (1996) and Yano
\& Gouda (1998a) extended their model to the case with
non-vanishing
$\dot{r}$.
They found that
the relation between $n$ and $\gamma$ is
$\gamma=[3h(n+3)]/[2+h(n+3)]$, where,
$h\equiv
-a\left<\dot{x}\right>/\dot{a}x$, which is a ratio of a peculiar
velocity to the Hubble expansion. With this result,
$\gamma$ can vary from 0 to 2 for
$n=1$ (Harrison-Zel'dovich spectrum) and $0\leq h \leq 1$ ($h=1$
corresponds to the Davis-Peebles solution).
  Since we do not know the stability of those solutions,
  in order to find which value of $\gamma$ is most likely,
we should study the dynamics of density fluctuations in other
methods, e.g. $N$-body simulation. Several groups in fact
showed that a power-law behavior in two-point correlation
function is obtained by $N$-body simulation with appropriate
primordial density fluctuations (Miyoshi \& Kihara (1975), Efstathiou (1979),
Aarseth, Gott III, \& Turner (1979), Frenk, White, \& Davis (1983),
Davis
et al. (1985), Jing (1998)).

The question is whether those power-law behaviors mean that we
have a fractal structure in the present Universe.
Peebles (1985) and Couchman \& Peebles (1998)
showed how to proceed with a high resolution analysis in the $N$-body
simulation using a kind of renormalization method.  They have
used Davis-Peebles solution as a scaling relation. Without such
an ansatz, we do not know whether usual $N$-body simulation is
suitable to discuss the formation of a fractal structure. With
the present state of computers, it may not be possible to obtain
high enough resolution to analyze a fractal structure.

As for a fractal structure in the Universe, one may ask another question.
Did the Universe not have any non-trivial structure such as a fractal
in the initial density fluctuations?
In the conventional approaches, initial density perturbations
are usually assumed to be given by a power-law (or a power-law-like) 
spectrum with random Gaussian phase.
Although such initial conditions may
provide the presently observed nonlinear scale-free structure via
nonlinear dynamics, no one has shown whether such a structure is
fractal or not, and if yes, what kind of fractal structure comes
out.
To provide a fractal structure in the present Universe, we may
adopt an alternative scenario, in which primordial density fluctuations
have already a fractal-like structure in the beginning.  Note that a
background spacetime is assumed to be a smooth universe, which is
described by the Einstein-de Sitter universe, but not a fractal
universe.  The properties of  an initial fractal may be preserved during
the evolution of the Universe,  then nonlinear fractal structure will be
formed. In fact, De Gouveia Dal  Pino et al. (1995) reported that the
temperature fluctuation of CMBR has a  fractal relation, and recently,
Pando \& Fang (1998) and Feng \& Fang  (2000) also reported that
non-Gaussianity was detected in the distribution  of Ly$\alpha$ forest
lines in the QSO absorption spectra.  In this  scenario, several natural
questions may arise. How does such a primordial fractal perturbation
evolve into nonlinear regime?  Will any properties of the initial fractal
be preserved during the evolution of the Universe, or not? If not, what
kind of nonlinear structure will come out at present? Is there any
fundamental difference in the structure formation process between a
conventional density perturbation and the present fractal one? In order
to answer those questions, we study the time evolution of the initial
density fluctuations with a fractal structure in Einstein de-Sitter
universe.


Since we are interested in a fractal structure, a quite high
resolution is required in our calculation.
As we discussed, $N$-body simulation may not have enough resolution
in the present state of computer development, unless we develop some 
skillful method.
So, in this paper,
we consider only a very simple toy model, which is a
one-dimensional (1-D) sheet model, in order to get some
insight into the questions raised in the above.
To set up primordial fractal density perturbations, we distribute
  $N$ sheets initially
by some systematic rule, i.e. we apply a Cantor set or
random Cantor-type set (see below).
Mathematically, in order to construct a Cantor set, the procedure
must be repeated an infinite number of times, but it is not practically 
possible to set up such
initial data. We therefore stop the procedure at a certain point, i.e.
the initial set is given by several times removing line
segments with a given ratio (Falconer (1990)). This could be
justified because an infinite scale-free structure never
exists in the real Universe. In order to construct the
initial density perturbations, we set that the remaining segments
have small positive density perturbations, while the removed ones
correspond to small negative ones. Since we study a 1-D sheet-model,
the motion of each sheet is described by an analytic solution
(Zel'dovich (1970)), which guarantees enough resolution to
analyze a fractal structure.

In $\S 2$, we present our formalism and initial setting.
As for the initial data, we consider three cases:
  regular Cantor set, random Cantor-type
set, and random white noise. Comparing those time evolutions,
we show our
results in $\S 3$.
In $\S 4$, we focus particularly on the phase space.
The conclusion and
discussion follow in $\S 5$.

\section{FORMALISM AND INITIAL DATA}
\subsection{Dynamical Equations}

In order to study the structure formation of the expanding Universe,
there are so far three approaches:
$N$-body simulation, the Eulerian perturbation approach and the
Lagrangian one.
Although the final answer for structure formation would be
obtained by $N$-body simulation, it may not be possible to so far
answer the questions about a fractal structure.
  As for the perturbation
approaches, these are just an approximation and will break down
in nonlinear regime, although the Lagrangian approach would be
better if we are interested in the density perturbations.
This is just because
a density fluctuation $\delta$ and a peculiar  velocity ${\bf v}$ are perturbed
quantities in the Eulerian approach (Peebles (1980)), while
displacement of particles is assumed to be small in
the Lagrangian approach (Zel'dovich
(1970), Bouchet (1992, 1995), Coles \& Lucchin (1995),
Catelan (1995)).
Its first order
solution is the so-called Zel'dovich approximation (Zel'dovich
(1970)).  The Lagrangian
approach is confirmed to be better than the Eulerian
approach by comparison of these results in several
cases (Munshi, Sahni, \& Starobinsky (1994),
Sahni \& Shandarin (1996), Yoshisato, Matsubara, \& Morikawa (1998)).

The perturbation variable ${\bf S}$ in the Lagrangian approach describes
a displacement of dust particles from a uniform
distribution and is defined  as:
\begin{eqnarray}
{\bf x} &=& {\bf q} + {\bf S} ({\it q}, t) \;,
\label{L-perturbation}\\ {\bf x} &:& {\rm Eulerian \; comoving
\; coordinate} \;, \nonumber \\ {\bf q} &:& {\rm Lagrangian \;
comoving \; coordinate} \;. \nonumber
\end{eqnarray}
The density fluctuation is given by the Jacobian $J$ as
\begin{eqnarray}
\delta({\bf q},t) & = & \frac{1-J}{J}
\end{eqnarray}
where $J  \equiv  {\rm det} \left ( \partial {\bf x}/\partial {\bf q}
\right )$.

Even though the Lagrangian approach is better, it is still an approximation
and then is not suitable to discuss a highly nonlinear structure
formation. However, there is one exceptional case.  If the distribution
is plane symmetric, the system is one-dimensional, and then the
Zel'dovich approximation turns out to be an exact solution.
Hereafter, we discuss only one-dimensional problem, which  Lagrangian
perturbation is given by 
\begin{eqnarray}
x = q + S ({\it q}, t) 
\end{eqnarray}
where $x$ and $q$ are the one-dimensional Eulerian and  Lagrangian
comoving coordinates, respectively.
For Einstein de-Sitter model,
the  solution is given by  (Gouda \& Nakamura (1989), Bouchet et al. (1995)):
\begin{equation}
S(q,t) = a(t) S_1(q)
+ a(t)^{-3/2} S_2(q) \;,
\label{sol-of-ZA}
\end{equation}
where the scale factor $a$ changes as $a=(t/t_0)^{2/3}$.
Then we find the position and its velocity of a dust particle at a
scale factor $a(t)$ with respect to the Lagrangian coordinate $q$
as
\begin{eqnarray}
{x} (q,a) &=& q + a(t) S_1(q)
+ a(t)^{-3/2} S_2(q) \;,
\label{position}
\\
\tilde{v}(q,a) &=&  S_1(q)
-{3 \over 2}  a(t)^{-5/2} S_2(q),
\label{velocity}
\end{eqnarray}
where we have introduced a new peculiar velocity  $\tilde{v}\equiv\partial
x/\partial a$.
In what follows, we use a scale factor   $a(t)$ as a time coordinate instead of
the physical time
$t$.

Although the Zel'dovich solution is exact,
  in studying a formation of nonlinear structure,  a serious
problem will soon arise. As a
density  fluctuation  grows, we find a shell crossing.
For a realistic matter fluid, a pressure may prevent such a singularity from
forming.
Then the solution will no longer describe the evolution of
perturbations after a shell crossing.
However, we may have another choice.
If instead of a usual matter fluid, we have a collisionless
particle such as some dark matter, we can go beyond the shell
crossing.  The particles, which are described by plane parallel sheets,
will pass through each other without collision.
Then after this crossing, we rediscover that the Zel'dovich
solution is again exact. Therefore, we have a series of exact
solutions, which is almost analytic
(Gouda \& Nakamura (1989), Yano \& Gouda (1998b)).
We shall call it the 1-D sheet model.

To be more precise, when a
crossing by two sheets has occurred, those two sheets
exchange their numbering as follows.
Suppose there are two sheets $q_1, q_2 (q_1< q_2)$ with the
Eulerian coordinates $x(q_1, a), x(q_2, a)~ (x(q_1, a) < x(q_2, a))$
  and  with velocities
$\tilde{v}(q_1, a), \tilde{v}(q_2, a)$, respectively.
Assume those sheets
cross over at
$a_{\rm cross}$, i.e.
$x(q_1, a_{\rm cross})= x(q_2, a_{\rm cross})$ with $\tilde{v}(q_1,
a_{\rm cross}) > \tilde{v}(q_2, a_{\rm cross})$.
For the evolution   after a shell crossing ($a>a_{\rm cross}$),
exchanging their numbering ($q_1 \rightleftarrow ~ q_2$) as
\begin{eqnarray}
x(q_1, a_{\rm cross})
&\rightarrow &x(q_2, a_{\rm cross}), ~~~~~\tilde{v}(q_1, a_{\rm cross})
~\rightarrow ~\tilde{v}(q_2, a_{\rm cross}) \;, \nonumber
\\
x(q_2, a_{\rm cross})
&\rightarrow &x(q_1, a_{\rm cross}), ~~~~~
\tilde{v}(q_2, a_{\rm cross})~\rightarrow
~\tilde{v}(q_1, a_{\rm cross}) \;,
\end{eqnarray}
we find again a natural ordering between the Lagrangian and Eulerian
coordinates, i.e.
$x(q_1, a) < x(q_2, a)$ for $q_1<q_2$.
By this exchange, we obtain a new distribution of
sheets (${x} (q,a), \tilde{v}(q,a)$) just after a shell
crossing. Using this distribution as an initial data, we find 
a next exact time evolution of the system by Zel'dovich solution.
In order to fix the initial data, 
 we have to determine
$S_1(q)$ and $S_2(q)$ in (\ref{sol-of-ZA}) for
the given distribution  (${x} (q,a), \tilde{v}(q,a)$). 
From
(\ref{position},\ref{velocity}) we find the solution is
\begin{eqnarray}
S_1(q) &=& \frac{3}{5a} (x-q) + \frac{2}{5} \tilde{v} \;,\nonumber \\
S_2(q) &=& \frac{2}{5} a^{3/2} \left \{ (x-q) - a \tilde{v}
\right \} \;.
\label{S_1S_2}
\end{eqnarray}
\\
  The new exact
solution (\ref{sol-of-ZA}) 	with (\ref{S_1S_2}) is valid until we
encounter next shell crossing.
another two sheets will cross over.

We repeat this prescription every time when we
encounter a shell crossing.  As a result, we obtain 
 a series of the Zel'dovich's
exact solutions, which is regarded as an analytic solution for
the 1-D collisionless sheet model. Note that this prescription is still
valid in multi-stream region. Using this
prescription, Gouda
\& Nakamura (1989) investigated a time evolution for the density
perturbations with a   scale-free initial
power spectrum. They showed that the power spectrum
will approach some characteristic value
independent of the power index of the initial spectrum.
This characteristic power index -1
is predicted by a catastrophe theory.
Recently, Yano \& Gouda (1998b) investigated a
time evolution of the density
perturbations for an initial power
spectrum with a cutoff. In this case, they found that
a self-similarity  in all scales is no longer valid. The
  spectrum is classified  into five ranges by its power index. Some
spectra
  coincide with the above one for some scale ranges, but 
 another power index, which is independent of the initial
power spectrum,   appears 
 just beyond the cut-off scale. In $\S 5$, we show  our result
comparing with their result.

Since this 1-D sheet model is powerful enough to see the fine
structure, we shall use this model to analyze a time evolution of
the primordial fractal density perturbations.

\subsection{Setting Up Initial Data}
Since we are interested in  initial  density fluctuations with a fractal
distribution, we have to construct such an initial data.
For the sake of simplicity, we apply
a Cantor set, or a random Cantor-type set (see below),
in our construction.  A Cantor set is given by the following
procedure.  We first divide a line with the length $L$ by some integer
$n_D$ and then remove one  line segment at the center.
If $n_D$ is an even integer, not dividing a line by $n_D$, we
just remove a line segment with the length
$L/n_D$ from the center of the line. We then repeat this procedure for
the remained line segments.  In
mathematical definition, the removal  procedure  must be
repeated infinite times.  However we believe  that a fractal structure, even
if it exists in the Universe,  is not a mathematical one but its 
self-similarity
may end at some scale.  Then we also stop our procedure after a
finite number of repetitions. We regard the remaining line segment as a 
region of
positive density fluctuations ($\delta_+$ region),
while the removed part corresponds to a region of
negative density fluctuations
($\delta_-$ region).
$\delta_+ (>0)$ and $\delta_-(<0)$ are chosen to be uniform, i.e.
both $\delta_+$ and $\delta_-$ are some constants in all regions
such that
$\delta_+ \simeq |\delta_-| \simeq 10^{-3}$.
Although this is very artificial and the realistic perturbations may depend
on each scale just like conventional density perturbations,
we analyze only this  simplest case here (see Discussion).
  In addition to the
Cantor set, in order to see the universality of our results, we also consider a
random Cantor-type set as well as  a distribution constructed by a white
noise. We describe in more detail how to construct the initial data for
each case.

\subsubsection{Regular Cantor set}
In this case, we consider  seven initial data.
Each data is constructed by removing central line segments
  with a fixed ratio ($1/n_D~(n_D=3, 6, 8,10,12, 15, 20)$) from
the remained parts.
  Each
density fluctuation has a different fractal dimension given by
\begin{equation}
D_{0} = \frac{\log 2}{\log[2n_D/(n_D-1)]} .\;
\end{equation}
We assume that  $\delta_+ \sim |\delta_-|$.
With this ansatz, we have fixed the repetition number of  removal
procedure ($N_R$).
Although  the number $N_R$ can be different for each initial data,
in order to keep  the same resolution for each model, (i.e. for the ratio  of
the smallest line segment
$\ell$ to that of the whole region $L$ (our calculation space) to be
almost same), we set $N_R= 5 \sim 7$.
We also fix   $\delta_+ =10^{-3}$, which determines the negative density
perturbation $\delta_-$ such that  the mean of fluctuations must vanish
  (see Table 1).

\tabcolsep=10mm
\begin{center}
\begin{tabular}{ccccc}
$n_D$ & $D_{0}$ & $N_R$ & $\ell/L$ & $\delta_-$\\ \hline
3 & 0.631& 5 & 1/243 &$-0.152\times 10^{-3}$
\\
6 &  0.792& 6 & 1/191.10 &$- 0.504\times 10^{-3}$\\
8 & 0.838 & 7 & 1/325.95 &$-0.647\times 10^{-3}$\\
10 & 0.868 & 7 & 1/267.62 &$-0.917\times 10^{-3}$\\
12 & 0.888 & 7 & 1/235.36 &$-1.19\times 10^{-3}$\\
15 & 0.909 & 7 & 1/207.47 &$-1.61\times 10^{-3}$\\
20 & 0.931 & 7 & 1/183.29&$-2.31\times 10^{-3}$ \\
\end{tabular}
\end{center}

\baselineskip=18pt
\begin{flushleft}
\noindent
Table 1 : \\The number of division $n_D$, its fractal dimension $D_{0}$,
the repetition number of removal procedure $N_R$, the resolution (the ratio 
of the
length of the shortest line segment to that of the whole system $\ell/L$), and
each negative density perturbation
$\delta_-$.
\end{flushleft}

\baselineskip=18pt

We show one example of initial data for $n_D=10$ in Fig. 1.
We put $2^{17}$ sheets in
our calculation space $L$.
By use
of a box-counting method (the box length $x$ is chosen from
$2^{-16} L (\sim 1.5 \times 10^{-5} L) $ to $10^{-1} L$), we have checked
the fractal dimension of initial fluctuations.
We
calculate the number $N(x)$ of boxes with a length $x$ in which
$\delta_+$-segments exist. The result is shown in Fig. 2.
We find a power law in the $x$-$N(x)$ relation for the range of
$10^{-3} L
\le x \le 10^{-1} L$. From this curve, we estimate the fractal
dimension as 0.868, which is the same value as that of the present
Cantor set.
Below  $x=10^{-3} L$,  we have a small deviation from a power law
relation, which corresponds to the limit of the resolution of our
present model.

Since the Cantor set is quite systematically constructed, one may
wonder that our  results strongly depend on such a special setting and may not
be  universal.
To answer such a question, we also analyze two different initial data
settings: one is a random Cantor-type set and the other is just a white
noise.

\subsubsection{Random Cantor-type set}
The random Cantor-type set is defined so that the division number $n_D$ is
fixed as that of a regular Cantor set, while its removal positions of line
segments are determined  by a random number.  According to the box-counting 
method,  we find
that the fractal dimension of initial fluctuations is almost the same
as that of a regular Cantor set for
the same $n_D$.  In this paper, we
analyze 3 random Cantor-type set models: two models with
$n_D=10$ and one with  $n_D=12$. One initial
distribution for $n_D=10$ is shown in Fig. 3.

\subsubsection{White Noise}
In order to see whether or not our primordial fractal fluctuations will
play an important role in a structure formation, in particular a
formation of fractal structure in a nonlinear regime,
we also study a time evolution of primordial fluctuations with white noise.
The distribution of initial density fluctuations is given
by a random number between $-10^{-3} \leq \delta \leq 10^{-3}$.
We analyze 2 models, for one of which the initial data is shown in Fig. 4.

\section{TIME EVOLUTION OF PRIMORDIAL FRACTAL DENSITY PERTURBATIONS}
In order to see how structures are formed,  we have to describe the
distribution by the Eulerian coordinate. It is convenient to
compare the distribution at each time by use of the comoving
coordinate $x$.
We set the initial scale factor
$a_0=1$.
Since our initial density fluctuation is $\sim 10^{-3}$, we find a first
shell crossing at
$a=a_{\rm cross} \simeq10^3$.
We perform our calculation  until $a=(2 \sim 3) \times 10^4$.

In order to see the detail to resolve a fractal structure, we put
$2^{17}$ sheets in our calculation scale $L$.
Using a  box-counting method, i.e, counting the number of boxes  which
contain the region with density perturbation $\delta$
larger than
$1$, we determine the fractal dimension of
the nonlinear structures.  The size of the box  ranges from
$2^{-16} L$ to $10^{-1} L$

We shall present our results for three types of  initial data in order.

\subsection{Regular Cantor Set}

First, we show the results for the regular Cantor set with $n_D=10$.
The time evolution is depicted in Fig 5.
Because we set
$\delta_+ \simeq |\delta_-| \simeq 10^{-3}$, a nonlinear structure
appears at $a \simeq 5 \times 10^2$ (Fig. 5(a)).
Before a shell
crossing, the pattern of density fluctuations remains similar to that
of initial distributions, although each separation is going to change
through the gravitational interaction.
We then find a shell crossing at $a=a_{\rm cross} \sim 10^{3}$
(Fig. 5(b)).
After the shell crossing,
  the trace of initial Cantor set gradually disappears, because of the 
exchange of the shells by a crossing
(Fig. 5(b)-(f)).
Although many sheets cross  each other,
the peculiar velocities do not vanish immediately because of
collisionless sheets, and then  the pattern of nonlinear structure
will change continuously.
After sufficient evolution of the  nonlinear structure, we find a
self-similarity in the structure, which seems to be fractal.
In fact, enlarging some regions, we find similar density distributions
(Fig. 6).

In order to judge whether such a structure is fractal or not, we
  use  a  box-counting method, which gives
a fractal dimension $D_F$.
Before a shell  crossing, we find that the dimension  $D_F$ decreases
in time from the initial value  $D_0=0.868$,  but the error in  the
estimation increases with time (Fig. 7(a)).
This is because  although the
pattern of initial density fluctuations remains even in a nonlinear stage
before a shell crossing,
change of each separation breaks the initial fractal distribution.
Then, the initial fractal distribution seems to disappear.

However, after a shell crossing, the fractal dimension starts to  increase
again and the error in the estimation becomes much smaller.
The fractal structure seems to recover.
More surprisingly,
the dimension $D_F$  approaches some constant ($D_{\rm asym}\sim
0.9$) after $a
\simeq 1.5\times 10^4$, which is a little bit different from the initial
fractal dimension $D_0 =$0.868 (Fig. 7.(b)). In fact,
  $D_F = 0.889 \pm 0.009$ at $a=1.5\times 10^4$ and
  $D_F = 0.890 \pm 0.002$ at $a=2\times 10^4$.
Although  $D_0 =0.868$ is out of the error bar of $D_F$, the difference is very
small. However we will see that $D_F$ is really independent of the
initial fractal dimension $D_0$ later in the case of different initial
distributions (see Fig. 10).

We also calculate the two-point spatial correlation function of nonlinear
structures. If the structure is fractal, we have a relation between
a fractal dimension and a power index of
  the correlation function
(Falconer (1990)).  The correlation function here is evaluated for the
nonlinear
regions where the density fluctuation $\delta$ becomes larger than $1$.
At
$a=10^3$, just after a first shell crossing, the correlation function
shows a rapid oscillation between a positive and a negative values
because of the periodic pattern of the Cantor set.
After a shell crossing, the pattern of Cantor set disappears,
and  then such an oscillation also vanishes.
After enough time passes, i.e. when the stable fractal structure is
found, the correlation function becomes also stable (Fig. 8).
The function is positive for the distance of  $x \le 5 \times 10^{-2}
L$, beyond which it becomes negative because  a shell
crossing does not yet occur for such a scale and  the largest
nonlinear structure is about $5
\times 10^{-2} L$. The trace of initial
Cantor set still remains.
If the structure is fractal,  the correlation
function must show a  power-law behavior.
In fact, from Fig. 8, we find a power
law relation in the range of
$10^{-4}L<x<10^{-2}L$ as $\xi \propto x^{-\gamma}$ with
$\gamma=0.130
\pm 0.005$.
When the fractal dimension  is
$D_F$, and the correlation function is given  as
\begin{equation}
\xi(x) \propto x^{-(1-D_F)} \;,
\end{equation}
then we find $\gamma =1-D_F$.  The evaluated  fractal dimension from
two point correlation  is
$D_F=0.870
\pm 0.005$ at $a=10000$, which is close to the value obtained by a
  box-counting
$D_F = 0.862 \pm 0.009$.  We show the time evolution of two "fractal" 
dimensions
$D_F$ obtained from two point correlation and by a box-counting
method in Fig. 9.  From  Fig. 9, we confirm that the power index
$\gamma$ is well correlated to the  fractal dimension $D_F$.

We find a fractal structure after a shell crossing.  The fractal dimension
is close to the fractal dimension of the initial density distributions.
Does this dimension reflect that of the initial distributions?
If this is so, why does it disappear once around a first shell crossing
time and recover at very late time?
In order to answer these question,
we have looked for other initial conditions with different
$n_D$s, i.e. $n_D = 3,6,8,12,15,20$.
The fractal dimensions $D_0$ of the initial density distributions are
given in Table 1.
As for the time evolution, we find similar behaviour for all models.
The evolution of "fractal"
dimension $D_F$ is  shown  in Fig. 10. Surprisingly, for all models, all
  $D_F$
approach about $0.9$ at
$a=(2 \sim 3)
\times 10^4$, which is the end of our calculation.
To confirm our result, we have checked the size-number ($x$-$N(x)$)
relation by a  box-counting method, which shows almost a straight line
as in Fig. 2.
Since the initial dimensions of primordial fluctuations
were different, we would conclude that the fractal
dimension obtained after nonlinear evolution is universal within our
numerical accuracy.

\subsection{Random Cantor-type Set}
Since the Cantor set is highly systematically constructed,
one may ask whether the present result strongly  depends on such a
special
initial setting.
  Is the universal dimension of nonlinear fractal structures due to the
primordial density fluctuations  defined by the regular Cantor set?
In order to answer this question, we shall analyze a different model
with randomness, which we call a random Cantor-type  set defined in
the previous subsection.

We analyze 3 models: 2 models with
$n_D=10$ (model 1 and model 2) and  1 model with $n_D=12$.
We find that just as in the case of regular Cantor sets, the fractal
dimension for nonlinear structures always approaches about
$0.9$ (Fig. 11).
We have also checked the size-number ($x$-$N(x)$) relation
in a  box-counting method, finding the same result as in the case of a 
regular Cantor set.
Because we  remove a line segment at a  random position, we usually
expect that the smallest segment is smaller than that of the regular
Cantor set as we show in the previous subsection.
As a result,  the stable fractal structure
will be formed later compared with the case with a regular Cantor set.
In fact, in the case of $n_D=12$, we find the stable dimension ($0.9$)
around
$a
\simeq 2.3 \times 10^4$ (Fig.11(c)).

\subsection{White noise case}
Then one may have another question.
Is the universal fractal dimension obtained above via
nonlinear dynamics and independent of the  initial distribution?
In order to answer this question, we also analyze a model with  white
noise fluctuation.
We analyze two models.  Both models do not show up
the above universal fractal dimension, although we find some different
stable asymptotic dimension ($\sim 0.7$) (Fig. 12).
The error in estimating the dimension is larger than that of the Cantor set
model, and the box-counting   shows some deviation from the
power law relation (Fig. 13).
Hence, it would not be a fractal.
We will discuss this in out Discussion.

\section{ANALYSIS OF PHASE SPACE}
Although the analysis by a box-counting suggests that a nonlinear fractal
structure with a universal dimension appears from primordial fractal
fluctuations, we may get more information from a detailed study of the
nonlinear structures obtained.
For this purpose, we analyze our result in a phase space.

We show the time evolution of the structures for the regular Cantor set with
$n_D=10$  in a phase space. Initially, the sheets
distribution in a phase space is given by a notched curve because
$\delta_+$ and $\delta_-$ are constant (Fig. 14).
If we enlarge the pictures, we find a similar notched curve  because of the 
present initial setting. These notches reflect a self-similarity in the 
initial
Cantor set.
  This behavior does not
change before a shell crossing (Figs. 14 (a), (b)).
Only the slopes of the line segments become steeper because of
the concentration of sheets.
After a shell crossing, the curve in a phase space shows
very complicated behavior.
Two sheets exchanged by a shell crossing are decelerated by the mutual
gravitational interaction, and then the curve will swirl (Gouda \& Nakamura
1989).  As the structure evolves, some
vortices are combined and form a larger vortex (Fig. 14).
When the "fractal" dimension becomes stable around 0.9, we find that the
large vortex consists of some similar small vortices. These small
vortices also consist of similar but much smaller vortices  (Fig.  15).
This discrete self-similarity in a phase space is found in all models.
As the structure evolves, some
vortices are combined and form a larger vortex (Fig. 14).
Although the initial fractal distribution seems to disappear, some trace
remains in the phase space.

One may wonder why  the centers of the vortices appear in
$\tilde{v} >0 $ region ( or $\tilde{v} <0 $ region).
The formation of a vortex in a phase space can be easily understood in the case
of a single wave ($S_1= \epsilon \sin q$) (see Gouda \& Nakamura (1989)).
In that case, the velocity at the center of the vortex vanishes.
Then the appearance of the vortex in $\tilde{v} >0 $ region
( or $\tilde{v} >0 $ region) seems inconsistent.
If we pursue each particle motion, nothing strange happens.
They move without swirling except at $\tilde{v} =0 $ point as shown in Fig. 16.

As was the case with a regular Cantor-type set,
we also find similar results in the phase space (Fig. 17).
However,  a discrete
self-similarity
  is hard to find,
although the larger vortices contain  smaller vortices as in the case of 
the regular Cantor-type set.
In the case with white noise fluctuation,
we cannot find any hierarchical vortex structure in a phase space (Fig. 18).

\section{CONCLUSIONS AND DISCUSSION}
We have studied  the nonlinear evolution of primordial
fractal fluctuations by using a 1-D sheet model.
We have analyzed 7 models
  with initial fluctuations constructed by a regular Cantor set, 3
models  with initial fluctuations constructed by a random
Cantor-type set, and 2 models with white noise fluctuations.
For all models except for the case with white noise, we find a kind
of attractor with a universal fractal dimension ($\sim 0.9$) as the
fluctuations evolve into nonlinear regime.
In the case with white noise fluctuations, the estimated
dimension becomes stable  around
$0.7$, but the error in the estimation  is larger than  the
other cases and the power-law behavior in a box-counting is
also not completely fitted.  Then, it may not contain a fractal
structure.
 From the phase space analysis, we find
a hierarchical structure, that is,
  the large vortex consists of some
similar small vortices, and such small vortices again consist
of similar but much smaller vortices. In particular, we find a
discrete self-similarity for the model with a regular Cantor set.

Why is the fractal dimension close to
$0.9$?
  Is it really universal?
Is the present fractal structure  really an attractor?
Although we need more analysis to
answer this question, we have some hints
in previous work.
Gouda and Nakamura studied the present 1-D sheet model for the
initial power law spectrum. They found two types of generic
singularities when we have a shell crossing (Gouda \& Nakamura
1988, 1989). When a first shell crossing appears, the relation
between Eulerian  and Lagrangian coordinates must be
\begin{equation}
x=q_c +\beta (q-q_c)^3 + \cdots,
\end{equation}
while that after a shell crossing turns out to be
\begin{equation}
x=q_c +\beta (q-q_c)^2 + \cdots.
\end{equation}
Following Arnold's classification, the former and latter cases are
classified  into  A3 and A2, respectively.
A3 is structurally unstable and may appear transiently
in the expanding Universe.
A2 is structurally stable and appears  universally for the
initial power-law spectrum.
The latter case  gives
\begin{eqnarray}
\delta_k &=& \int \delta (x) e^{ikx} dx \nonumber \\
& \propto & \int (\beta x)^{-1/2}e^{ikx} dx \nonumber \\
& =& (\beta k)^{-1/2} \int
\eta^{-1/2} e^{i\eta} d\eta,
\end{eqnarray}
i.e. $P(k) \sim k^{-1}$.
This predicts $\gamma = 0$, i.e. $D_F=1$, which
  is rather close to
our ``universal" dimension 0.9.
Although one may wonder that
these are essentially the same,
we have another result which suggests that
there seems to exist a new type of
stable phase.  Recently, Yano and Gouda
analyzed a more realistic case, i.e. the initial power law
spectrum with a cut-off and
found 5 characteristic regions in Fourier space (Yano \&
Gouda 1998a, b).
The regime 1 is the linear one  and then it is
just an initial power spectrum. In the regime 2, they found
$P(k) \sim k^{-1}$, which is the single-caustic regime
(Gouda \& Nakamura (1989)).
The regime 3 is called the multi-caustic regime, in which the power
spectrum depends on the initial power-law index.
Beyond the cut-off scale, two regimes appear, one gives
$P(k) \sim k^{-1}$ (regime 5), which may correspond to
A2 type stable solution.
In the intermediate wave number $k$   between the regime 3
and regime 5, they find $k^\nu$, which $\nu$ is independent
of initial power  index and close to 1, but a little less.
  They called
it the virialized regime. This seems to be a new transient region,
which may appear in  some specific initial conditions.
We would conjecture that the fractal structure with a universal dimension 0.9
corresponds to this virialized regime (regime 4)
  and the dimension 0.7 found in the case with white noise
would be the regime 3.
By reanalyzing the Yano-Gauda model in the case of $k=0$, we
have confirmed that  $\nu =\sim 0.9$.
We also find  a small tail with
index 0.7 in the size-number relation in the
Cantor set model with $n_D =15, 20$ (Fig. 19).

This conjecture is also supported by the analysis for a
self-gravitating 1-D sheet model
without the background expansion of the Universe
(Tsuchiya, Konishi \& Gouda 1994).
They found  two time scales; one is a micro relaxation time
($t_{\rm micro} = N t_c$) and the other is a global relaxation time
($t_{\rm global} = 4 \times 10^4 N t_c$), where $t_c
=\sqrt{L/4\pi G N m}$ is a crossing time.  After $t_{\rm micro} $,
some equilibrium state is reached by exchanging particle energy,
  but the global relaxation is not
achieved, i.e. the partition function is not yet described
by an equilibrium state such as an ergodic state
  (Tsuchiya, Konishi \& Gouda 1994). In the present
model, we can speculate that the fractal structure is obtained
after this micro relaxation time but before the global relaxation
time. In fact, if we estimate the above time scales in the present
models, we find that
$t_{\rm micro}$ corresponds to
$a=5\times 10^3$, while $t_{\rm global}$ corresponds to
$a=5\times 10^6$.
The time when we find a stable fractal structure ($a=(1 \sim 3)
\times 10^4$) is  between those two time scales.
If this speculation is true,
our fractal structure is temporal.
In the future of the Universe, it will evolve into
more relaxed and ergodic state.

Since we analyze the simplest case, we have to extend our
analysis  to more generic cases.
First, we should study different types of fractal in order to check
whether the present results are universal for any fractal distributions or not.
  Secondly, we need to analyze the case with scale-dependent
fluctuations. In the present analysis,
we set $\delta_+=$ constant and $\delta_-=$ constant.
In the realistic case, there must be a scale dependence
to the fluctuations. In the conventional  perturbations, we usually
assume a power-law spectrum with some cutoff.
Even if the primordial fluctuations contain a fractal structure,
their amplitude may depend on the scale.
Its dependence may change the present results.
In particular, in the present model,
the scenario of structure formation
could be different from either TOP-DOWN or BOTTOM-UP
for some scale-dependence.
The primordial fractal fluctuations will evolve directly into a
hierarchical nonlinear structure.
But, it will definitely depend on the scale dependence of the
fluctuations. Secondly, we need to extend the present analysis to
other cosmological models, i.e.
the open Universe  model and $\Lambda \neq 0$ flat Universe
model.
For the 1-D sheet  model, the solutions are still exact, and
the growth and  decay rates in
  these models are different from those in the Einstein-de Sitter
Universe  model.
We expect that  the structure formation  after a shell crossing
is not the same as that in the present cosmological   model, and
  then the fractal dimension would be different.

For more realistic cases, we must study either the 2-D or 3-D model.
Since the   Zel'dovich solution is no longer exact,
we have to explore a new method.
In order to preserve  a high  resolution, we may
develop a kind of  renormalization method in $N$-body simulation
as Couchman \& Peebles (1998).

Finally, it would also be interesting to look for the origin
of such a primordial fractal density perturbation.
The inflationary scenario may provide the origin of primordial
fluctuations. One may wonder whether such a fractal primordial
fluctuation is expected in some inflationary models.
If we have more than two scalar fields, then the system is
not integrable and may show a chaotic behavior or a fractal
property (Easther \& Maeda (1999)).
Such a model might show up a kind of fractal density perturbation.

\acknowledgements

We would like to thank N. Gouda, P. Haines, O. Iguchi, T. Kurokawa,
V. Lukash, M. Morikawa, A. Nakamichi, Y. Sota, T.
Yano, and A. Yoshisato for many useful discussions.
Part of this work was done while
KM was participating the program, ``Structure Formation in the
Universe'', at the Newton Institute, University of Cambridge.
KM is grateful to the Newton Institute for their hospitality.
  Our numerical
computation  was carried out by  Yukawa Institute Computer Faculty.
  This work was supported partially by a Grant-in-Aid for
Scientific Research Fund of the Ministry of Education, Science
and Culture (Specially Promoted Research No. 08102010),  and by the
Waseda University Grant  for Special Research Projects.

\newpage

\begin{flushleft}
Figure Captions:
\end{flushleft}

\noindent
Fig.1. The initial fractal density fluctuation  for the regular Cantor set with
$n_D=10$. Enlarging the picture, we find the same pattern up to the
repetition number of  removal procedure $N_R=7$.

\noindent
Fig. 2. The size-number ($x$-$N(x)$) relation in a  box-counting method for
the model in Fig. 1. The dotted line shows
$N(x)
\propto  x^{-0.868}$.
We find that the similarity exists in the range between $10^{-3} L$ and
$10^{-1}L$.  Below  $x=10^{-3}L$, we have a deviation from the power law
relation, which corresponds to our resolution limit ($\ell /L \sim 1/267.62$).

\noindent
Fig. 3.(a) The initial fractal density fluctuation  for the
  random Cantor-type  set with
$n_D=10$. The random Cantor-type set is defined so that the division number 
$n_D$ is fixed as that of a regular Cantor set, while its removal positions 
of line segments are determined by a random number. This is called
the  model 1 in the text.
Enlarging the picture, we find the similar pattern up to the
repetition number of  removal procedure $N_R=7$.  The fractal dimension of
this initial fluctuation is
$0.868$, which  is the same as that of the regular Cantor set with $n_D=10$.

\noindent
(b) Another initial fractal density fluctuation  set up by the same
prescription as
 (a) (model 2).

\noindent
Fig. 4.(a) The initial  density fluctuation created by a
  white noise. This is called
the  model 1 in the text.

\noindent
(b) Another initial  density fluctuation  set up by the same
prescription as
 (a) (model 2).

\noindent
Fig. 5.(a)  Time evolution of density fluctuation for the case with
$n_D=10$. This  snap shot is at
$a=  5
\times 10^2$, when the positive
fluctuation $\delta_+$ grows just to a
nonlinear scale ($\delta_+ =1$).

\noindent
(b) The same as (a) but at $a\sim a_{\rm cross}\sim 10^3$, which is just after
a first shell crossing.

\noindent
(c) The same as (a) but at $a=6 \times 10^3$.

\noindent
(d) The same as (a) but at $a=10^4$.

\noindent
(e) The same as (a) but at $a=1.5 \times 10^4$.

\noindent
(f) The same as (a) but at $a=2 \times 10^4$.

\noindent
Fig. 6. Hierarchical nonlinear structure at $a=10^4$. Enlarging the picture, we
find the self-similar structure.

\noindent
Fig. 7.(a) The evolution of the "fractal" dimension $D_F$ of nonlinear 
structure
for the case of $n_D=10$. Before a shell  crossing ($a < a_{\rm cross}
\sim 10^3$).

\noindent
(b) The same as (a) but after a shell crossing ($a \ge  a_{\rm cross}
\sim 10^3$).

\noindent
Fig. 8. Two-point correlation function of nonlinear structures at
$a=10^4$.
The dotted line is $\xi \propto x^{-0.130}$.

\noindent
Fig. 9. The time evolution of  the "fractal" dimensions
$D_F$ obtained from two point correlation and by a
box-counting method
  The circles denote $D_F$  by two-point correlation ($D_F=1-\gamma$), while
the squares are those produced by a box-counting. Two methods to determine  the
fractal dimension agree well each other.

\noindent
Fig. 10.(a) The time evolution of  the "fractal" dimension  $D_F$ of
nonlinear structure for $n_D=3$.

\noindent
(b) The same as (a) but $n_D=6$.

\noindent
(c) The same as (a) but $n_D=8$.

\noindent
(d) The same as (a) but $n_D=12$.

\noindent
(e) The same as (a) but $n_D=15$.

\noindent
(f) The same as (a) but $n_D=20$.

\noindent
Fig. 11.(a) The evolution of the "fractal" dimension $D_F$  of nonlinear
structure in the case of the random Cantor-type initial fluctuations for
$n_D=10$ (model 1).  The
dimension approaches about
$0.9$ at $a=2
\times 10^4$.

\noindent
(b) The same as (a) ($n_D=10$, model 2). The
dimension approaches about
$0.9$ at $a=2.8 \times 10^4$.

\noindent
(c) The same as (a) ($n_D=12$). The
dimension approaches about
$0.9$ after $a=2.3 \times 10^4$.

\noindent
Fig. 12.(a) The evolution of the "fractal" dimension $D_F$  of nonlinear
structure in the case with white noise fluctuations (model 1). The
dimension approaches about $0.7$.

\noindent
(b) The same as (a) but of model 2.
We also find the same stable dimension $0.7$.

\noindent
Fig. 13. The size-number ($x$-$N(x)$) relations in a  box-counting method 
for the models in Fig. 12.  There is some small deviation from a power-law 
behaviour.

\noindent
Fig.14.(a) Structure Formation in phase space ($n_D=10$). At $a=1$. The
distribution draws a self-similar graph, because initial density fluctuation
was given by a regular Cantor set.

\noindent
(b) The same as (a) but at $a=10^3$. At immediately after shell crossing,
the sheets just began production of vortices.

\noindent
(c) The same as (a) but at $a=3 \times 10^3$. The self similarity of
vortices exists during four steps.

\noindent
(d) The same as (a) but at $a=5 \times 10^3$. The self similarity of
vortices exists during five steps.

\noindent
(e)  The same as (a) but at $a=10^4$. Two
of $L$ scale structure was begun to combined.

\noindent
(f) The same as (a) but at $a=1.5 \times 10^4$.

\noindent
(g) The same as (a) but at $a=2 \times 10^4$.

\noindent
Fig. 15. A sheet distribution at $a=2 \times 10^4$ in
phase space ($n_D=10$).
The dotted squared regions are enlarged in the next figures shown by arrows.
We find a discrete self similarity, i.e. a large vortex consists
  of some similar small vortices, and those small
vortices also consist of similar but much smaller vortices.

\noindent
Fig. 16. The orbits of 7 particles in phase space ($n_D=10$, 6000 $\le a 
\le 20000$). Each particle did not swirl, however distribution of particles 
become to set of vortices.

\noindent
Fig. 17. A sheet distribution at $a=30000$ in phase space
  for the model with a random
Cantor-type set
  ($n_D=12$). In this figure, the larger vortices contain   smaller vortices as
the case of regular Cantor set, but a discrete
self-similarity
  is hard to be found.

\noindent
Fig. 18. A sheet distribution  $a=30000$ in
phase space for the model with white noise
fluctuations (model 1). The vortices were widely spread and a nest of
vortices is rarely found.

\noindent
Fig. 19. The size-number ($x$-$N(x)$) relation in a box-counting method for 
the regular Cantor set with $n_D=15$ at $a=30000$. The dotted line denotes 
$N(x) \propto  x^{-0.9}$, while the dotted-dash line is $N(x) \propto 
x^{-0.7}$.

\end{document}